\documentclass[nofootinbib,prl,amsmath,amssymb,twocolumn,floatfix]{revtex4-2}

\usepackage{graphicx}% Include figure files
\usepackage{dcolumn}% Align table columns on decimal point
\usepackage{bm}% bold math
\usepackage{siunitx}
\renewcommand{\epsilon}{\varepsilon}
\newcommand{\M}{\mathcal{M}}
\newcommand{\F}{\mathcal{F}}

\begin{document}

\preprint{APS/123-QED}

\title{Energy estimates for wave amplification in quasiperiodic Fibonacci time-modulated media}

\author{I. Ruchiev and B. Davies$^*$}
\affiliation{Mathematics Institute, University of Warwick, Coventry, CV4~7AL, United Kingdom \\
$^*$bryn.davies@warwick.ac.uk}

\date{\today}

\begin{abstract}
Fibonacci time quasicrystals can be approximated by temporal supercells to reveal a fractal collection of $k$ gaps, in which wave energy is amplified exponentially. These estimates are validated by the observation of ``super'' $k$ gaps that are independent of the duration of the temporal supercell. This approach predicts the regions of parametric amplification and provides accurate estimates of the energy growth rate.
%
% We show that the energy... The first observation of super k gaps: momentum gaps that are independent of the duration of the temporal supercell approximation. Gives accurate estimate of energy growth rate (as well as whether or not it grows). First time Fibonacci modulation in time has been studied.
\end{abstract}

\maketitle

\textit{Introduction} --- There has been longstanding interest in understanding wave propagation in media whose properties vary in time \cite{morgenthaler1958velocity, holberg1966parametric, oliner1961wave, simon1960action}. This has been renewed by the recent development of space--time metamaterials, which offer unprecedented freedom to manipulate waves \cite{galiffi2022photonics, yin2022floquet}. Examples of exciting phenomena include non-reciprocal transmission \cite{simon1960action, oliner1961wave}, Fresnel drag \cite{huidobro2019fresnel}, parametric amplification \cite{cullen1958travelling, raiford1974degenerate}, exceptional points \cite{nikzamir2022achieve, rouhi2020exceptional}, topological localisation \cite{rechtsman2013photonic, fleury2016floquet} and frequency conversion through the coupling of harmonic modes \cite{koutserimpas2023multiharmonic, zurita2009reflection, hiltunen2024coupled}. As well as posing challenging theoretical questions, recent experimental breakthroughs mean these systems are now a reality \cite{tirole2023double, peng2016experimental}.

Time modulation, even of spatially homogeneous materials, has deep implications for wave propagation. For example, a temporal interface leads to a reflected (time-reversed) wave \cite{morgenthaler1958velocity, Fink1997TimeReversed}. When the modulation is periodic in time, systems often display parametric intervals of stability and instability \cite{morgenthaler1958velocity, holberg1966parametric, zurita2009reflection, pierrat2025causality, koutserimpas2018electromagnetic}. The dependence on wavenumber yields alternating intervals of stable propagation and amplification in so-called \emph{$k$ gaps}  \cite{zurita2009reflection}. As well as periodic modulation, random \cite{carminati2021universal, pierrat2025causality} and quasiperiodic temporal modulation \cite{koufidis2024enhanced, naweed2025unconventional, he2025experimental} have been studied. In both cases, amplification phenomena analogous to $k$ gap behaviour have been observed.

Quasicrystals are exciting phases of matter that exhibit long-range order without translational symmetry \cite{shechtman1984metallic}. Systems with spatially quasiperiodic modulation often exhibit exotic spectral properties, such as fractal and Cantor-like spectra \cite{Simon1982}, unexpected symmetries in reciprocal space \cite{shechtman1984metallic}, unique topological properties \cite{liu2022topological, putley2024mixing, agazzi2014colored} and transitions between universally localised or propagating modes \cite{jitomirskaya1999metal, morison2022order}. One of the most widely studied examples of (spatial) quasicrystals is the \emph{Fibonacci quasicrystal} \cite{jagannathan2021fibonacci}. Defined based on a tiling rule, its self-similar properties lend themselves to concise analysis \cite{kohmoto1983localization}.

The unusual spectral properties of quasicrystals have inspired their use in wave control and metamaterial applications, such as topological waveguides \cite{kraus2012topological, bandres2016topological, davies2022symmetry, beli2025interface} and graded frequency filters \cite{davies2023graded}; see Section~11 of \cite{davies2025roadmap} for a summary. These breakthroughs have had to overcome the fact that the exotic spectra of quasicrystals are exceptionally difficult to compute and they have relied on methods that apply Bloch's theorem through supercell and superspace approximation methods \cite{Shubin1978, davies2025convergence, damanik2015spectral}, as well as associated homogenisation techniques \cite{wellander2018two}.

\begin{figure}
    \centering
    \includegraphics[width=\linewidth]{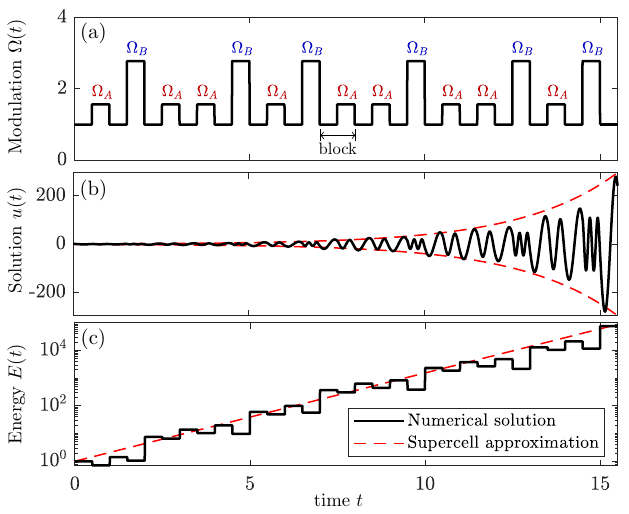}
    \caption{Piecewise-constant modulation according to a Fibonacci rule. (a) The coefficient $\Omega(t)$ alternates between $\Omega_0$ and either $\Omega_A$ or $\Omega_B$, depending on the corresponding letter in the Fibonacci word. (b) For certain parameter values (here, $k=3.2$) the solution grows exponentially. (c) The (exponential) rate of energy growth can be estimated using a supercell approximation.} \label{fig1}
\end{figure}

In this Letter, we reveal how Fibonacci time quasicrystals (FTQs) exhibit parametric regions of stability and amplification. Our main contribution is showing how temporal supercell approximations can be used to model FTQs. We demonstrate, for the first time, the existence of \emph{super $k$ gaps}: momentum gaps that exist for any sufficiently long supercell approximations. This phenomenon shows that supercells give bona fide predictions of amplification in FTQs. Using supercells and super $k$ gaps, we reveal the first concrete evidence of a fractal pattern of $k$ gaps in FTQs. A previous study of FTQs observed amplification of the reflected field \cite{naweed2025unconventional}; our methods reveal that this amplification occurs in parametric regions resembling the exotic spectral properties observed for spatial Fibonacci quasicrystals \cite{jagannathan2021fibonacci}. Finally, we can use supercells to forecast the average rate of energy amplification with arbitrarily good accuracy. This improves on existing estimates for random media \cite{carminati2021universal, pierrat2025causality}, demonstrating that the optimal way to characterise quasiperiodic modulation is through periodic approximation, rather than averaging.

% These exotic properties are indicative of the fact that spectra of quasicrystals are exceptionally difficult to compute. In practice, the most popular method is to approximate the quasicrystal as being periodic, using a \emph{supercell}. We have convergence... \cite{davies2025convergence, Shubin1978, damanik2015spectral}

\textit{Problem setting} --- 
We study the propagation of an electromagnetic wave in a medium with dielectric function $\varepsilon(t)$ that is modulated in time. We restrict our study to one spatial dimension and suppose that the material is homogenous and non-magnetic such that the system can be modelled by the linear wave equation
\begin{equation}\label{WE}
	\frac{\partial^2 u}{\partial x^2}(x,t) - \frac{\varepsilon(t)}{c^2}\frac{\partial^2 u}{\partial t^2} (x,t) = 0.
\end{equation}
% together with appropriate boundary conditions, to be discussed below. 
Taking the spatial Fourier transform leads to an ODE in reciprocal space parametrised by the wavenumber $k$:
\begin{equation}\label{ODE}
		\frac{\partial^2 \hat{u}(k,t)}{\partial t^2} + \Omega^2\hat{u}(k,t) = 0,
\end{equation}
where we have defined $\Omega =\Omega(k,t) := c^2k^2/\varepsilon(t)^2$.

In this study of FTQs, we will focus on the canonical case where the modulation $\Omega(t)$ is piecewise constant in time, modulated through a series of temporal interfaces (as in Fig.~\ref{fig1}(a)). Periodic \cite{lustig2018topological, biancalana2007dynamics, koutserimpas2018electromagnetic}, random \cite{carminati2021universal} and quasiperiodic \cite{naweed2025unconventional} versions of this system have been studied previously, as well as piecewise constant space--time modulation \cite{lurie2007introduction, milton2017field}. This setting is convenient since, on any section where $\Omega(t)=\Omega_j$ is constant, the solution to \eqref{ODE} can be written as
\begin{equation}\label{expressionD}
	\hat{u}_j(t) =  \alpha_je^{-i\Omega_j t} + \beta_je^{i\Omega_j t}
\end{equation}
for some constants $\alpha_j,\beta_j$ which need to be chosen such that the solution is continuous across the jumps in $\Omega(t)$. To achieve a realisation of a FTQ, we consider the case where $\Omega(t)$ alternates between a background value $\Omega_0$ and one of two values $\Omega_A$ and $\Omega_B$, which are selected according to a Fibonacci tiling rule \cite{jagannathan2021fibonacci} (this is depicted in Fig.~\ref{fig1}(a)). This tiling rule gives a pattern of `$A$'s and `$B$'s that is defined sequentially, where each new word in the sequence is formed by combining the previous two. If the first two terms are $\mathcal{F}_0 = B$ and $\mathcal{F}_1 = A$, then we form subsequent words according to
\begin{equation} \label{Fibonacci}
	\mathcal{F}_n = \mathcal{F}_{n-1} \cup \mathcal{F}_{n-2}.
\end{equation}
This gives $\mathcal{F}_2 = AB$, $\mathcal{F}_3 = ABA$, $\mathcal{F}_4 = ABAAB$ and so on. As $n\to \infty$, the ratio of the number of `$A$'s to `$B$'s in the sequence $\mathcal{F}_n$ converges to the golden ratio, $\varphi=(1+\sqrt{5})/2$.

\textit{Transfer matrices and supercells} --- 
We wish to understand under what conditions the interactions between waves in the FTQ and its quasiperiodic temporal modulation lead to stable propagation or amplitude growth. In particular, we want to elucidate the dependence on the wavenumber $k$ and understand the analogue of the $k$ gaps observed for periodic temporal modulation \cite{holberg1966parametric, zurita2009reflection}. 

We use the methods and analytic framework from \cite{carminati2021universal} and will exploit the general form \eqref{expressionD} of solutions to \eqref{ODE}. Successive values of the constants $\alpha_j$ and $\beta_j$ in \eqref{expressionD} can be related by a transfer matrix $M=M(\Omega_j)$:
\begin{equation} \label{eq:TM}
	\begin{bmatrix}
		\alpha_{j+1} \\ \beta_{j+1}
	\end{bmatrix}
	= M(\Omega_j)
	\begin{bmatrix}
		\alpha_j \\ \beta_j
	\end{bmatrix}.
\end{equation}
For details, including explicit expressions for the entries of $M(\Omega_j)$, see \cite{carminati2021universal}. Propagation on time intervals spanning many blocks of modulated $\Omega(t)$ can be described by products of many of the matrices ${M}(\Omega_{j})$.

% Then, each `block' in the modulation of $\Omega(t)$, as labelled in Fig.~\ref{fig1}(a), can be described by the matrix $\mathcal{M}_{j}=M(\Omega_{j})M(\Omega_0)$, where $j=A,B$. Further, on time intervals spanning many blocks of modulated $\Omega(t)$, we can consider products of many of the matrices $\mathcal{M}_{j}$.

\begin{figure}
    \centering
    \includegraphics[width=0.95\linewidth,trim=0.2cm 0 0.7cm 0,clip]{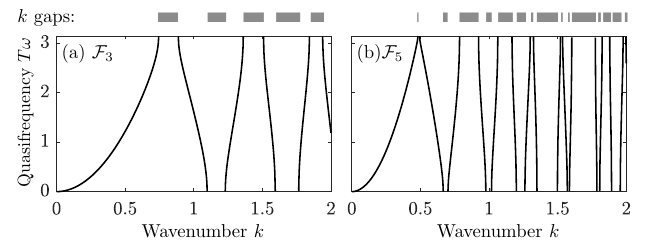}
    \caption{Dispersion curves for the temporal supercells given by the Fibonacci words (a) $\F_3=ABA$ and (b) $\F_5=ABAABABA$. The $k$ gaps are shaded above each plot.}
    \label{fig4}
\end{figure}

\begin{figure*}
    \centering
    \includegraphics[width=\linewidth]{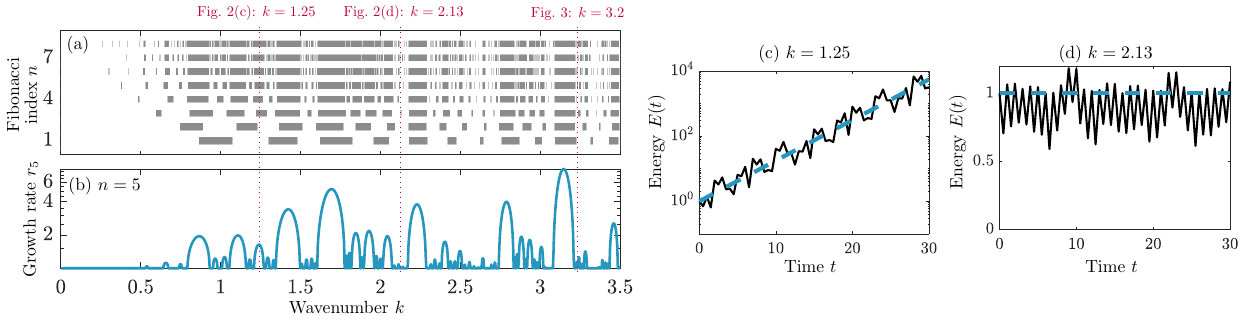}
    \caption{Supercell approximations can be used to predict stability. Here, (a) The pattern of $k$ gaps for successive supercell approximations. The $k$ gaps are shaded for successive Fibonacci unit cells, indexed by $n$. (b) The growth rate, estimated using the $\mathcal{F}_5$ supercell. The energy $E(t)$ for (c) $k=1.25 $ and (d) $k=2.13$. In both cases, the data are plotted alongside the slope predicted by the $\mathcal{F}_5$ supercell approximation.}
    \label{fig2}
\end{figure*}

An example of a solution in the FTQ is shown in Fig~\ref{fig1}(b), where the parameters have been chosen such that the amplitude grows exponentially. The parameter values $c=1$, $\epsilon_A=0.8$ and $\epsilon_B=0.6$ are taken, which we will use for all subsequent numerical examples presented in this Letter. Understanding for which wavenumbers amplification occurs is a challenging problem (related to the notorious spectral problems for spatial Fibonacci quasicrystals \cite{Simon1982}). To characterise this, we will approximate the quasicrystal as being periodic, using a temporal \emph{supercell}. That is, we take sections of the quasiperiodic modulation as unit cells for periodic modulation (allowing us to use Bloch's theorem). 

The key to using the supercell method to study quasicrystals is to choose a sequence of successively larger supercells to get increasingly good approximations. Here, we elect to take the unit cells that are generated by the tiling rule \eqref{Fibonacci}. For each unit cell in this sequence, we have an associated transfer matrix
\begin{equation}
    \mathcal{M}_{\mathcal{F}_n} = \prod_{j\in\mathcal{F}_n} M(\Omega_j)M(\Omega_0).
\end{equation}
We can use Bloch's theorem to compute dispersion curves, as in \cite{zurita2009reflection}. In particular, we have the dispersion relation
\begin{equation} \label{eq:dispersion}
    \cos(T\omega)=\frac{1}{2} \mathrm{tr}[\M_{\mathcal{F}_n}(k)],
\end{equation}
where $\omega$ is a quasifrequency that is the temporal Bloch parameter. Deriving \eqref{eq:dispersion} uses the fact that $\det(\M)=1$, which is equivalent to the condition that $|\alpha_j|^2-|\beta_j|^2$ is conserved for all $j$ \cite{carminati2021universal}.

The crucial quantity that governs whether a periodic supercell solution grows or not is the maximum eigenvalue of the associated transfer matrix $\mathcal{M}$. If $(\lambda_k,\mathbf{v}_k)$ are the eigenpairs of $\M$, then we can write the initial conditions as $(\alpha_0,\beta_0)^\top=A\mathbf{v}_1+B\mathbf{v}_2$ and then
\begin{equation} \label{eig_growth}
    \begin{bmatrix}
		\alpha_{jT} \\ \beta_{jT}
	\end{bmatrix}
    = A \lambda_1^j \mathbf{v}_1+B \lambda_2^j \mathbf{v}_2,
\end{equation}
where $T$ is the duration of the temporal supercell. Thus, the solution will grow exponentially if one of the eigenvalues is bigger then 1, i.e. if $\max\{|\lambda_1|,|\lambda_2|\}>1$. Since $\mathrm{tr}[\M_{\mathcal{F}_n}]$ is the sum of the eigenvalues, this condition is equivalent to  the dispersion relation \eqref{eq:dispersion} having a real-valued solution for $k$ and $\omega$.

We can use the dispersion relation \eqref{eq:dispersion} to study the pattern of $k$ gaps the sequence of supercell approximations. Two examples of dispersion curves are shown in Fig.~\ref{fig4}, for the supercells $\F_3=ABA$ and $\F_5=ABAABABA$. Whenever $k$ is such that there is no real-valued solution, we say that $k$ is in a $k$ gap; these regions are shown above each plot in Fig~\ref{fig4}. The pattern of $k$ gaps for successive Fibonacci-generated supercell approximations is shown in Fig.~\ref{fig2}(a). As the duration of the supercell increases, the pattern of $k$ gaps becomes increasingly intricate, appearing to converge towards a fractal pattern, reminiscent of the fractal, Cantor-like spectral that are often seen for spatial Fibonacci quasicrystals \cite{jagannathan2021fibonacci}.

\textit{Super $k$ gaps} --- 
One of the striking features of Fig.~\ref{fig2}(a) is the occurrence of gaps that persist independent of the index $n$ and the duration of the supercell $\F_n$. This behaviour is reminiscent of the {super band gaps} that are observed when periodic approximations are applied to spatial quasicrystals \cite{davies2024super, morini2018waves}. Similarly, we can define \emph{super $k$ gaps} as values of $k$ which are in $k$ gaps of the supercells $\F_n$ for all $n\geq n_0$. 

The existence of super $k$ gaps can be verified by studying the traces of the matrices $\M_{\F_n}$. From the dispersion relation \eqref{eq:dispersion}, we can see that $k$ is in a $k$ gap if and only if $|\mathrm{tr}[\M_{\F_n}(k)]|>2$. A very convenient property of the tiling rule \eqref{Fibonacci} is that the traces of the associated transfer matrices can be related to one another. In particular, \cite{kohmoto1983localization} discovered that if $x_n=\mathrm{tr}[\M_{\F_n}]$, then 
\begin{equation} \label{eq:kohmoto}
    x_{n+1}=x_n x_{n-1}-x_{n-2}.
\end{equation}
This uses the fact that $\det(\M_{\F_n})=1$. This recursion relation can be used to show that if $k$ is in a $k$ gap for both $\F_N$ and $\F_{N+1}$, for some $N$, then it will be in a $k$ gaps for $\F_n$ for all $n\geq N$ \cite{SBGcomment}. Hence, any such values of $k$, which we can easily spot in Fig.~\ref{fig2}, must be in super $k$ gaps. The reason it is exciting to identify super $k$ gaps is that they are parameter regions where we can be confident that the supercell approximations are accurately predicting the instability.

\textit{Energy growth rate} --- 
As well as predicting stability and instability, supercell approximations can be used to predict the long-term behaviour of the wave's energy. In particular, we want to estimate the rate at which the energy grows when the system is unstable. We define the energy as
\begin{equation}
    E_j = |\alpha_j|^2 + |\beta_j|^2.
\end{equation}
This is constant on any interval where $\Omega(t)$ is constant. However, it may grow whenever $\Omega(t)$ has a jump, as seen in Fig~\ref{fig1}(c).

For any periodic modulation, \eqref{eig_growth} can be used to estimate the energy growth rate. In particular, if $\lambda:=\max\{|\lambda_1|,|\lambda_2|\}$, then we have that $\alpha_j,\beta_j\sim \lambda^{j/T}$ which implies that
\begin{equation} \label{eq:energy}
    E(t)\sim \exp\left( 2\frac{\log(\lambda)}{T} t \right),
\end{equation}
where $T$ is the duration of the supercell. Figs.~\ref{fig2}(c) and \ref{fig2}(d) show examples of energy profiles for parameters in a $k$ gap and parameters corresponding to stable propagation, respectively. In both cases, the estimate \eqref{eq:energy} based on the $\F_5$ supercell is shown, which gives a good prediction of the behaviour. The growth rate $\log(\lambda_{\F_5})/T$ is shown as a function of $k$ in Fig.~\ref{fig2}(b), which gives a measure of the `depth' of the super $k$ gaps.

\begin{figure}
    \centering
    \includegraphics[width=0.95\linewidth]{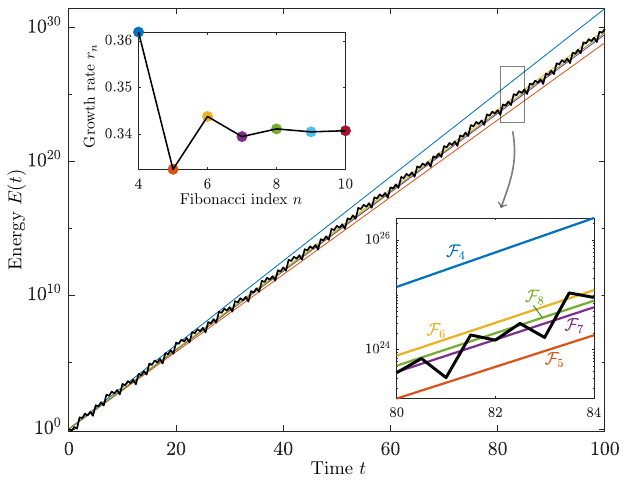}
    \caption{The supercell estimates for the energy growth rate converge as the duration of the unit cell is increased. Here, $k=3.2$ is in a super $k$ gap (see Fig.~\ref{fig2}(a)). The growth rate is estimated as $E(t)=\exp(r_n t)$ where the growth rate $r_n$ is based on the $\mathcal{F}_n$ supercell.  The convergence of the growth rates $r_n$ is illustrated inset.}
    \label{fig3}
\end{figure}

The accuracy of the estimate \eqref{eq:energy} for quasiperiodic modulation can be improved by using longer supercells. In Fig.~\ref{fig3}, we show the energy $E(t)$ for parameter values in a $k$ gap, along with growth rate estimates for several different Fibonacci unit cells $\F_n$. In the inset plot, we see that the growth rate estimate converges in the limit of large $n$. This shows that an arbitrarily good approximation of the average growth rate can be obtained by taking a sufficiently long supercell.

To see why the energy growth rate estimates converge, note that they are given by $r_n=\log(\lambda_{\F_n})/\mathbb{F}_n$, where $\mathbb{F}_n$ is the $n$th Fibonacci number (which is the duration of the $\F_n$ unit cell). Since $\det(\M_{\F_n})=1$, if we are in a $k$ gap, then we have that $x_n=\mathrm{tr}(\M_{\F_n})=\lambda_{\F_n}+1/\lambda_{\F_n}$. In a super $k$ gap, we expect $\lambda_{\F_n}$ to be very large when $n$ is large, so we can approximate $x_n\approx\lambda_{\F_n}$ in this regime. Recalling \eqref{eq:kohmoto}, when the values are large we have that $x_{n+1}\approx x_{n}x_{n-1}$, or equivalently $\log(x_{n+1})\approx\log(x_{n})+\log(x_{n-1})$. The limiting behaviour of this recursion relation can be characterised by taking the ansatz $\log(x_{n})=\rho^n$. This yields the characteristic polynomial $\rho^2=\rho+1$, which shows that $\log(x_{n})$ will grow according to $\log(x_{n})\sim\varphi^n$ when $n$ is large, where $\varphi=(1+\sqrt{5})/2$ is the golden ratio. It is also a well-known property of Fibonacci numbers that $\mathbb{F}_n\sim\varphi^n$ when $n$ is large. Putting this together, we have that the supercell growth rates converge
\begin{equation}
    r_n=\frac{\log(\lambda_{\F_n})}{\mathbb{F}_n} \sim \frac{\log(x_n)}{\mathbb{F}_n}\xrightarrow[n \to \infty]{}R,
\end{equation}
for finite $R$, which is the average growth rate of the energy of waves in the FTQ.

% It is worth noting that the analogous growth estimates for random modulation derived in \cite{carminati2021universal} are not applicable here. Explain... Intuition... This confirms the intuition that to predict the properties of quasicrystals accurately, it's better to approximate them as being periodic than as being random.

\textit{Conclusions} --- 
Using supercell approximations, we have revealed the intricate pattern of $k$ gaps exhibited by FTQs. This has analogues with the spectra of equivalent spatially modulated systems. The key difference is the exponential growth of solutions. Hence, the key new theoretical insight is that supercell approximations accurately predict the rate of energy growth. Further, the existence of super $k$ gaps guarantees that these approximations will converge, giving arbitrarily good predictions of the energy growth rate for sufficiently long temporal supercells.

% A direction that has not been explored here is whether supercell approximations can be used to characterise densities of states in FTQs. This is a 

% Something about density of states? Refer to \cite{ammari2025universal} and other papers

\begin{acknowledgments}
\textit{Acknowledgements ---} The authors gratefully acknowledge insightful discussions with Lorenzo Morini, Marc Martí Sabaté and T. V. Raziman during the development of this work. The code used to generate the numerical examples presented here is available for download \cite{ruchiev2025fibonacci}. 
\end{acknowledgments}

\bibliography{references}

\end{document}